\documentclass[aps,prl,twocolumn,groupedaddress,showpacs,superscriptaddress]{revtex4-1}
\usepackage{graphicx,amsmath}
\usepackage{amssymb}

\begin{document}
\title{Analytic continuation of the self-energy via Machine Learning techniques}

\author{Taegeun Song}
\affiliation{Department of Physics, Pohang University of Science and Technology, Pohang 37673, Republic of Korea}
\affiliation{The Abdus Salam International Centre for Theoretical Physics (ICTP), Strada Costiera 11, I-34151 Trieste, Italy}
\author{Roser Valent\'\i}
\affiliation{Institute of Theoretical Physics, Goethe University Frankfurt, Max-von-Laue-Strasse 1, 60438 Frankfurt am Main, 
Germany}
\author{Hunpyo Lee} \email{Email: hplee@kangwon.ac.kr}
\affiliation{School of Liberal Studies, Kangwon National University, Samcheok, 25913, Republic of Korea}
\date{\today}

\begin{abstract} 
We develop a novel analytic continuation method for self-energies on the Matsubara domain as computed by quantum Monte Carlo 
simulations within dynamical mean field theory (QMC+DMFT). Unlike a maximum entropy (maxEn) procedure employed for the last thirty 
years, our approach is based on a machine learning (ML) technique in combination with the iterative perturbative theory 
	impurity solver of the dynamical mean field theory self-consistent process (IPT+DMFT).
	The input and output training datasets for ML 
are simultaneously obtained from IPT+DMFT calculations on Matsubara and real frequency domains, respectively.
	The QMC+DMFT self-energy on real 
frequencies is determined from the -usually noisy- input QMC+DMFT self-energy
	on the Matsubara domain and the trained ML kernel. Our approach is 
	free from both, bias of ML 
training datasets and from fitting parameters present in the maxEn method.
We demonstrate the efficiency of the method on the testbed frustrated Hubbard model on the square lattice.

\end{abstract}

\maketitle

{\it Introduction.-} Electronic properties in strongly correlated systems have been intensively studied within the dynamical mean 
field theory (DMFT) 
approximation~\cite{Metzner1989,Georges1992,Georges1996,Kotliar2004,Held2007} by making use of 
powerful quantum Monte Carlo (QMC+DMFT) methods as impurity 
solvers~\cite{Rubtsov2005,Werner2006,Gull2011,Imada1998,Kotliar2006}. While QMC+DMFT has successfully accounted for
a variety of phases such as Fermi-liquid, non-Fermi-liquid and paramagnetic Mott insulator, to mention a few, the calculated impurity Green's 
functions  $G_{\sigma} (i\omega_n)$ (or self-energy $\Sigma_{\sigma} (i\omega_n)$) with spin index $\sigma$, retain numerical noise 
in the Matsubara domain ($i\omega_n$)~\cite{Rubtsov2005,Werner2006,Gull2011}. This noise creates unfortunately large uncertainties 
in the results of $G_{\sigma}(\omega)$ after performing analytic continuation to real frequencies $\omega$ with the methods 
presently at hand.

The most widely employed tool for analytic continuation of $G_{\sigma}(i\omega_n)$ is the maximum entropy (maxEn) 
method~\cite{Jarrell1996}. This technique is based on defining a goodness-of-fit functional $\chi^2$ and entropy $S$ associated with
the spectral function $A_{\sigma}(\omega) = − \frac{1}{\pi} \rm{Im} G_{\sigma} (\omega + i0^+)$. The optimized solution is then 
determined by minimizing $F(\alpha) = \chi^2 - \alpha^{-1}S$ where $\alpha$ is a parameter which controls the degree of 
regularization. However, the results of the maxEn analytic continuation are strongly dependent on $\alpha$~\cite{Wang2009}. 
Furthermore, for a direct comparison to experimental observations, analytic continuation of the self-energy  $\Sigma_{\sigma}^{\text 
{QMC+DMFT}} (i\omega_n)$ calculated within DMFT+QMC is required. The uncertainty in the analytic continuation of 
$\Sigma_{\sigma}^{\text {QMC+DMFT}} (i\omega_n)$ is much larger than that of $G_{\sigma}^{\text {QMC+DMFT}} (i\omega_n)$, due to the 
inversion problem of $G_{\sigma}^{\text {QMC+DMFT}}(i\omega_n)$ in the Dyson's equation~\cite{Wang2009}. Therefore, for analytic 
continuation from the Matsubara to real frequency domain the development of more reliable tools that are absent of fitting 
parameters is desirable.

\begin{figure}
\includegraphics[width=1.00\columnwidth]{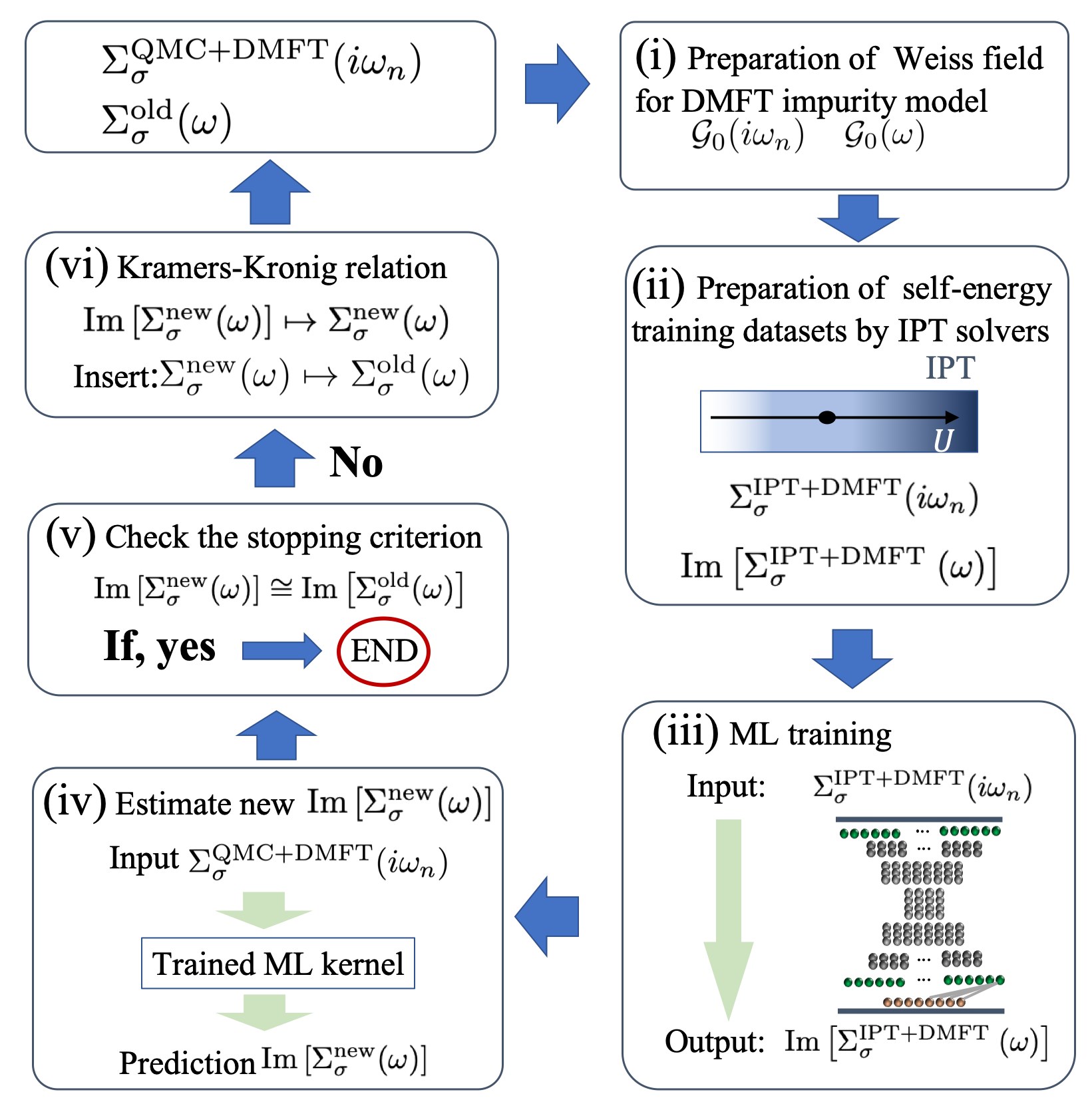}
\caption{\label{Fig1} (Color online) Schematic architecture of the analytic continuation in combination with the machine learning 
(AC+ML) method. 'IPT' and 'DMFT' denote iterative perturbation theory impurity solver and dynamical mean field theory approximation, 
respectively. The AC+MC simulation begins in (i)  with converged self-energy $\Sigma_{\sigma}^{\text {QMC+DMFT}} 
(i\omega_n)$ data and unknown self-energy $\Sigma_{\sigma}^{\text {old}} (\omega)$. Initially, $\Sigma_{\sigma}^{\text {old}} 
(\omega)$ are set to zero in all real frequencies. The input and output training datasets for machine learning (ML) are 
simultaneously determined by IPT+DMFT on, respectively, Matsubara and real frequency domains in (i) and (ii). The QMC+DMFT 
self-energy at real frequencies is computed by the trained ML parameters based on convolutional autoencoder tool (iii) and (noisy) 
input  $\Sigma_{\sigma}^{\text {QMC+DMFT}} (i\omega_n)$ in (iv). When the condition of $\text{Im} [\Sigma_{\sigma}^{\text{new}} 
(\omega)] \approx \text{Im} [\Sigma_{\sigma}^{\text{old}} (\omega)]$ in (v) is satisfied, the patterns of the AC+ML self-energy 
$\Sigma_{\sigma}^{\text {AC+ML}} (\omega)$ on real frequencies are fully rearranged, otherwise a new full self-energy is
computed by the Kramers-Kronig relation in (vi) and the complete process is repeated. The convergence of the AC+ML method is done 
within several iterations.}
\end{figure}

Recently, a procedure for analytic continuation of $G_{\sigma}(i\omega_n)$
to real frequencies was proposed by Yoon {\it et al.}~\cite{Yoon2018} on the
basis of a machine learning (ML) approach. The authors first generated arbitrary spectral functions $A_{\sigma}^{\text{training}} 
(\omega)$ out of approximately $10^5$ different configurations classified by number of peaks, their height and their position.
Since computing $G_{\sigma} (i\omega_n)$ for
a given $A_{\sigma}(\omega)$ is straightforward, for supervised ML 
$G_{\sigma}^{\text{training}} (i\omega_n)$ was calculated from:
\begin{equation}\label{Eq1}
G_{\sigma}^{\text{training}} (i\omega_n)=\int d\omega \frac{A_{\sigma}^{\text{training}} (\omega)} {i\omega_n-\omega}.
\end{equation}
Next, $G_{\sigma}^{\text{training}} (i\omega_n)$ and $A_{\sigma}^{\text{training}} (\omega)$ were employed as input and output 
training datasets in the ML process, respectively. $A_{\sigma}^{\text{ML}} (\omega)$ was then estimated by the trained parameters of 
the supervised ML where $G_{\sigma} (i\omega_n)$ are the input data. The advantage of this ML approach is the absence of a fitting 
parameter, unlike the case of the maxEn approach. On the other hand, the results are strongly biased by the configurations of 
the selected training datasets $A_{\sigma}^{\text{training}} (\omega)$.


In this Letter we suggest an alternative analytic continuation based on a combination
of a machine learning (AC+ML) approach with DMFT data extracted from iterative
perturbation theory~\cite{Kajueter1996,Tremblay2012,Fujiwara2003} (IPT+DMFT) where quantities are obtained in
Matsubara and real frequency domains simultaneously.
This combined method ensures the absence of a fitting 
parameter present in the maxEn method and systematically makes the ML training datasets without 
possible biasing originating from the selections of the training datasets. Moreover, our AC+ML 
approach directly performs the analytic continuation of the converged QMC+DMFT self-energy $\Sigma_{\sigma}^{\text{QMC+DMFT}} 
(i\omega_n)$ to real frequencies.  We present results for 
the frustrated Hubbard model on the square
lattice and demonstrate the efficiency of the method.

{\it Analytic continuation based on Machine learning.-} The idea of the AC+ML
method is the following; we take notice of the IPT+DMFT method to make training
datasets for the ML process.
With IPT+DMFT self-energies can be calculated in both
Matsubara and real frequencies at the same time. The method is
known to capture 
various phases such as Fermi-liquid, Mott insulator and the Mott transition, as
observed in correlated electronic systems~\cite{Kajueter1996,Tremblay2012,Fujiwara2003}. 
The ML kernel is then
constructed from unbiased input datasets $\Sigma_{\sigma}^{\text{IPT+DMFT}}
(i\omega_n)$ and output datasets 
$\Sigma_{\sigma}^{\text{IPT+DMFT}} (\omega)$, prepared by the IPT+DMFT self-consistent equation. The AC+ML self-energy 
$\Sigma_{\sigma}^{\text {AC+ML}} (\omega)$ in the real domain is predicted by
the trained ML parameters based on a convolutional autoencoder tool~\cite{LeCun2015, Goodfellow2016, Krizhevsky2012, tensorflow} and input
QMC+DMFT self-energy data $\Sigma_{\sigma}^{\text {QMC+DMFT}} (i\omega_n)$ on
the Matsubara domain.


 The DMFT self-consistent equation
 for the real frequency-dependent Green's function can be 
expressed as
\begin{equation}\label{Eq2}
G_{\sigma}(\omega) =\sum_k \frac{1}{\omega + i\delta - \epsilon_k + \mu - \Sigma_{\sigma} (\omega)}, 
\end{equation}
where $\delta$, $\mu$ and $\epsilon_k$ are the broadening factor, chemical potential and energy dispersion, 
respectively~\cite{Georges1996}. The DMFT self-consistent equation in
the Matsubara domain is identically given as 
Eq.~(\ref{Eq2}), where $\omega + i\delta$ is replaced by $i\omega_n$. 

In Fig.~\ref{Fig1} we display the workflow of the simulation procedure
of the AC+ML approach. We start with (i) the calculation of the DMFT Weiss fields  $\mathcal{G}_0(i\omega_n)$ and $\mathcal{G}_0(\omega)$
with IPT in order to compose the training datasets of ML. For that we consider Eq.~(\ref{Eq2}) and
the Dyson's equation, and include the unknown self-energy $\Sigma_{\sigma}^{\text{old}} (\omega)$ in real frequencies.
The unknown self-energy $\Sigma_{\sigma}^{\text{old}} (\omega)$ are initially set to zero for all
frequencies. We also multiply $\Sigma_{\sigma}^{\text{old}}
(\omega)$ by  $\gamma$ to make more configurations for ML, where $\gamma$ are
real numbers larger than 
one. When $\gamma \Sigma_{\sigma}^{\text{old}} (\omega)$ are employed in
Eq.~(\ref{Eq2}), more new configurations for ML are composed. As the number of values
$\gamma$ is increased, both accuracy of analytic continuation and computational
expense increase. In a next step (ii) IPT+DMFT self-energies $\Sigma_{\sigma}^{\text{IPT+DMFT}} (i\omega_n)$ and 
$\Sigma_{\sigma}^{\text{IPT+DMFT}} (\omega)$ with a 
few hundreds of different configurations are calculated by Eq.~(\ref{Eq3}) given below for input and output training datasets, 
respectively. In step (iii) the ML parameters are determined by the training input datasets of 
$\Sigma_{\sigma}^{\text{IPT+DMFT}} (i\omega_n)$ and output datasets of 
the imaginary-part of self-energy $\text{Im} [\Sigma_{\sigma}^{\text{IPT+DMFT}} (\omega)]$  via a one dimensional 
convolutional autoencoder in Tensorflow-gpu~\cite{LeCun2015, Goodfellow2016,
Krizhevsky2012, tensorflow}. After this, in step (iv) the new imaginary-part of
the self-energy $\text{Im} [\Sigma_{\sigma}^{\text{new}} (\omega)]$ 
is estimated by the trained ML 
parameters with input data $\Sigma_{\sigma}^{\text{QMC+DMFT}} (i\omega_n)$ in Matsubara frequencies. In a next step (v), if the 
condition of $\text{Im} [\Sigma_{\sigma}^{\text{new}} (\omega)] \approx
\text{Im} [\Sigma_{\sigma}^{\text{old}} (\omega)]$ is satisfied, the AC+ML
simulation is over. Otherwise, after a new full self-energy
$\Sigma_{\sigma}^{\text{new}} (\omega)$ is computed 
by the Kramers-Kronig relation, it is inserted into Eq.~(\ref{Eq2}) as $\Sigma_{\sigma}^{\text{old}} (\omega)$, and the AC+ML 
simulation is repeated again. In most cases the convergence of the AC+ML is
done within several iterations.

For the ML process of Fig.~\ref{Fig1}, as mentioned above, we use a convolutional autoencoder based on stochastic variational
Bayes~\cite{Bengio2014, Kingma2013, Song2019} with gradient-based optimization~\cite{Kingma2014}.  The architecture of our ML kernel 
consists of encoder, decoder, and fully connected layer. Compared to a 
previous architecture~\cite{Song2019}, we use additional channels for dealing with both real- and imaginary-parts of 
$\Sigma_{\sigma}^{\text {DMFT+QMC}} (i\omega_n)$ input datasets. We employ $i\omega_{n_{\text{max}}}$ input nodes, where 
$i\omega_{n_{\text{max}}}$ means the maximum number of Matsubara frequencies in $\Sigma_{\sigma}^{\text {DMFT+QMC}} (i\omega_n)$.
The output of the decoder is connected with a single fully connected layer which consists of $\omega_{\text{max}}$ nodes, where 
$\omega_{\text{max}}$ is the number of real frequencies in $\text{Im} [\Sigma_{\sigma}^{\text{new}} (\omega)]$.


The IPT+DMFT self-energy on the real domain in (ii) of Fig.~\ref{Fig1} is computed by
\begin{equation}\label{Eq3}
\Sigma_{\sigma} (\omega) = -U^2 \int d\nu d\nu' d\nu'' \frac{A^{\text{L}}(\nu,\nu',\nu'') + A^{\text{R}}(\nu,\nu',\nu'')}{\omega - 
\nu + \nu' - \nu''},
\end{equation}
where $A^{\text{L}}(\nu,\nu',\nu'')$ and $A^{\text{R}}(\nu,\nu',\nu'')$ denote 
$A^{-}(\nu)A^{+}(\nu')A^{-}(\nu'')$ and $A^{+}(\nu)A^{-}(\nu')A^{+}(\nu'')$, respectively. Here, $A^{+}(\nu)$ 
and $A^{-}(\nu)$ are $f(\nu)A(\nu)$ and $f(-\nu)A(\nu)$ respectively, where $f(\nu)$ is the Fermi function expressed in real frequencies $\nu$
and  $A(\nu)$ is the spectral function at frequency $\nu$.  
The sum rules of $\text{Im} [\Sigma_{\sigma}^{\text{new}} (\omega)]$ in (iv) of Fig.~\ref{Fig1} are also conserved by
\begin{equation}
\int d\omega \text{Im} [\Sigma_{\sigma}^{\text{new}} (\omega)] = -\pi U^2 <n_{-\sigma}>(1-<n_{-\sigma}>),
\end{equation}
where $U$ and $<n_{\sigma}>$ are the electronic interaction
and occupation of electrons,  
respectively, 
in the QMC+DMFT calculation. The IPT+DMFT self-energy on  the Matsubara domain is identically expressed as Eq.~(\ref{Eq3}), where $\omega + 
i\delta$ is only replaced into $i\omega_n$.


\begin{figure}
\includegraphics[width=1.0\columnwidth]{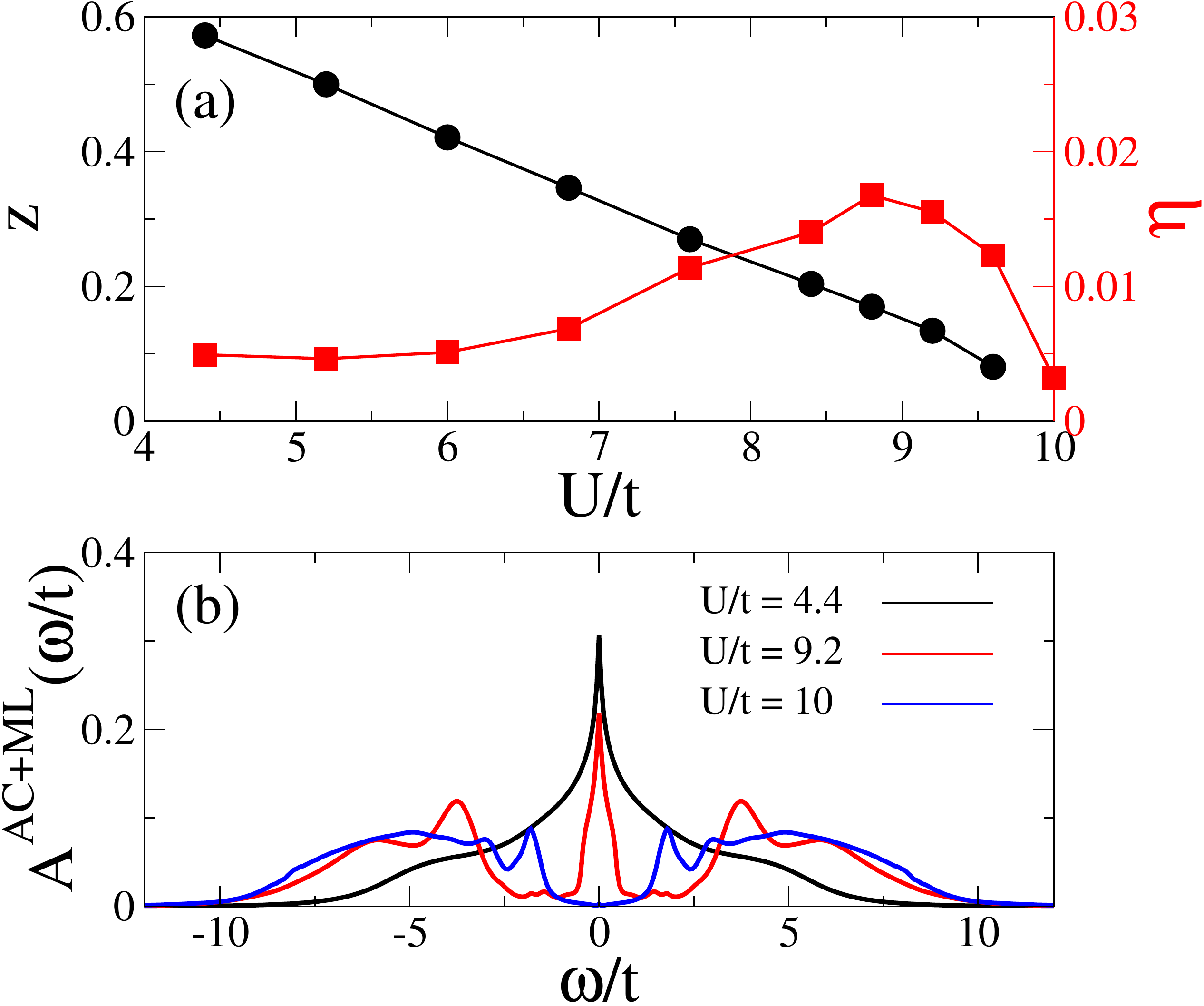}
\caption {\label{Fig2} (Color online) (a) (Left) Quasiparticle weight $Z$ and (Right) $\eta$ 
	as a function of $U/t$ for temperature $T/t=0.05$. $\eta$ is defined as $\eta = \frac{1}{\beta} \sum_{\omega_n} 
\vert (G_{\sigma}^{\text{QMC+DMFT}} (i\omega_n) - G_{\sigma}^{\text{AC+ML}} (i\omega_n) \vert$, where $G_{\sigma}^{\text{AC+ML}} 
(i\omega_n)$ are recovered by Eq.~(\ref{Eq1}). (b) Spectral function $A_{\sigma}^{\text{AC+ML}} (\omega)$ for several 
$U/t$. All results are obtained from the AC+ML procedure starting with the converged QMC+DMFT self-energy $\Sigma{\sigma}^{\text{QMC+DMFT}} 
(i\omega_n)$ on the square lattice at half filling with $t'/t=0.0$ and $\mu=0.0$.}
\end{figure}

{\it Results.-} In order to test the method, we  consider the Hubbard
Hamiltonian on the square lattice with nearest- and next-nearest-neighbor
hoppings
$t$ and $t'$ respectively in the QMC+DMFT calculations. Then $\epsilon_k=-2t (\cos (k_x)+\cos (k_y))-2t' (\cos (k_x+k_y) + \cos(k_x-k_y))$ and
the interaction part of the Hamiltonian $H_I$  is given
as 
\begin{equation}
H_I = U\sum_i (n_{\uparrow,i} - \frac{1}{2})(n_{\downarrow,i} - \frac{1}{2}),
\end{equation}
where $n_{\sigma,i}$ is the number operator at each spin $\sigma$ and site $i$.
We perform the AC+ML method with the converged QMC+DMFT
self-energy $\Sigma{\sigma}^{\text{QMC+DMFT}} (i\omega_n)$
where we set $\mu=0.0$. The temperature and the nearest-neighbor hopping strength employed for all QMC+DMFT simulations are $T/t=0.05$ and $t=1.0$,
respectively.
In order to make the ML training datasets in all IPT+DMFT simulations we
 set $\delta=0.03$ in Eq.~(\ref{Eq2}).

In Fig.~\ref{Fig2} (a)
we plot the quasiparticle weight $Z$ 
defined as  $Z=[1 - \frac{\partial\rm{Re} \Sigma_{\sigma}^{\text{AC+ML}} (\omega)}{\partial \omega}\vert_{\omega=0}]^{-1}$ 
and the parameter $\eta = \frac{1} {\beta} \sum_{\omega_n} \vert (G_{\sigma}^{\text{QMC+DMFT}} (i\omega_n) - G_{\sigma}^{\text{AC+ML}} (i\omega_n) 
\vert$ as a function of $U/t$ for the half-filled case first
for $t'/t=0.0$ and $\mu=0.0$. $\eta$ provides an estimate of the deviation of the AC+ML results
from the original data in the Matsubara domain.
 The values of $\gamma$ used in the 
IPT+DMFT self-consistent process to make ML training configurations are $1.0$, $1.2$, $1.4$, and $1.6$. 
Here $G_{\sigma}^{\text{AC+ML}} (i\omega_n)$ is recovered by
making use of Eq.~(\ref{Eq1}) via the spectral function.
We observe that $Z$  
decreases with increasing $U/t$, corroborating
many former results~\cite{Georges1996,Imada1998}. We identify the
Fermi liquid to Mott insulator 
transition around $U/t=9.8$. The parameter $\eta$ mostly increases
with increasing of $U/t$ in the Fermi liquid regions.
Note that  
the accuracy of the analytic continuation can be
improved with increasing number of $\gamma$ values.

Fig.~\ref{Fig2} (b) displays the
spectral function $A_{\sigma}^{\text{AC+ML}} (\omega)$ for several $U/t$. 
$A_{\sigma}^{\text{AC+ML}} (\omega)$ is determined by  
\begin{equation}
A_{\sigma}^{\text{AC+ML}} (\omega) = -\frac{1}{\pi} \text{Im} [G_{\sigma}^{\text{AC+ML}}(\omega)],
\end{equation} 
where $G_{\sigma}^{\text{AC+ML}}(\omega)$ has been computed from Eq.~(\ref{Eq2}) with $\Sigma_{\sigma}^{\text{AC+ML}} (\omega)$.
As we expect, the quasiparticle peak at the Fermi level $(\omega=0)$
decreases with increasing $U/t$ in the metallic regimes. The 
lower and upper Hubbard bands appear at large $U/t$ 
and the Mott gap is visible at $U/t=10.0$.

\begin{figure}
\includegraphics[width=1.0\columnwidth]{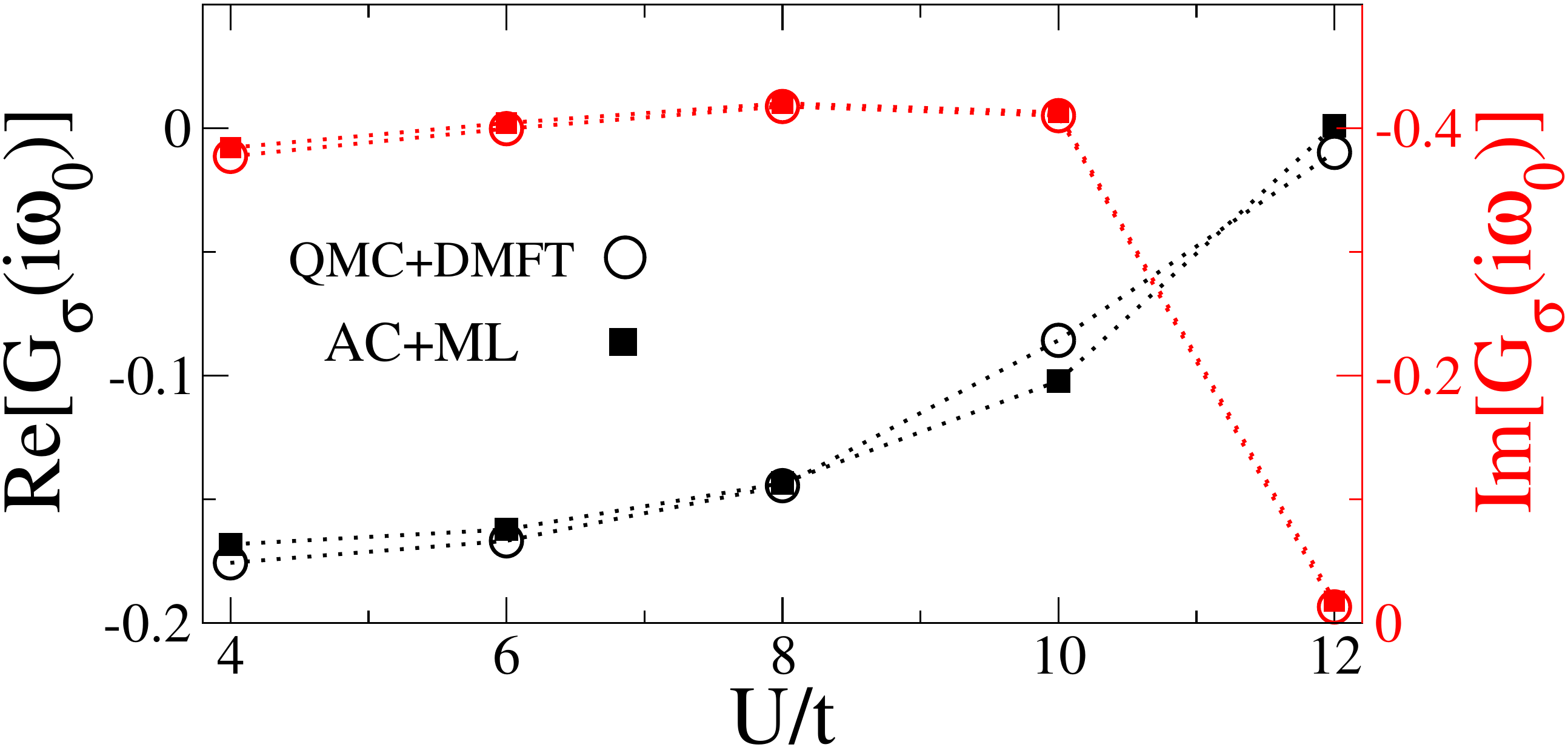}
\caption {\label{Fig3} (Color online) (Left) Real-part of the Green's function $\text{Re} [G_{\sigma}(i\omega_0)]$ and 
(Right) imaginary-part of the Green's function $\text{Im} [G_{\sigma}(i\omega_0)]$ at the first Matsubara frequency $i\omega_0$ as a 
function of $U/t$. The results are determined by the AC+ML method of the converged QMC+DMFT self-energy 
$\Sigma{\sigma}^{\text{QMC+DMFT}} (i\omega_n)$ for the fully frustrated Hubbard model with $t'/t=1.0$ and $\mu=0.0$ on the square 
lattice.}
\end{figure}

In the following we consider the  frustrated Hubbard model~\cite{Tocchio2013}
with $t'/t=1.0$ and $\mu=0.0$. The uncertainty of the 
analytic continuation in the frustrated system is much larger
than that in the half-filled particle-hole symmetric system with 
$t'/t=0.0$ because the real-part of the self-energy $\text{Re} [\Sigma_{\sigma}
(i\omega_n)]$ displays charge fluctuations in the QMC+DMFT calculations.
Here, we also tune $\mu$ to include more training datasets for charge
fluctuations in the IPT+DMFT 
self-consistent process. In order to confirm that our AC+ML approach is working well, in Fig.~\ref{Fig3} we plot real- and 
imaginary-parts of the Green's function at the first Matsubara frequency $i\omega_0$ as a function of $U/t$ in both AC+ML and original 
QMC+DMFT cases. We find that both results are in an excellent agreement.

\begin{figure}
\includegraphics[width=1.0\columnwidth]{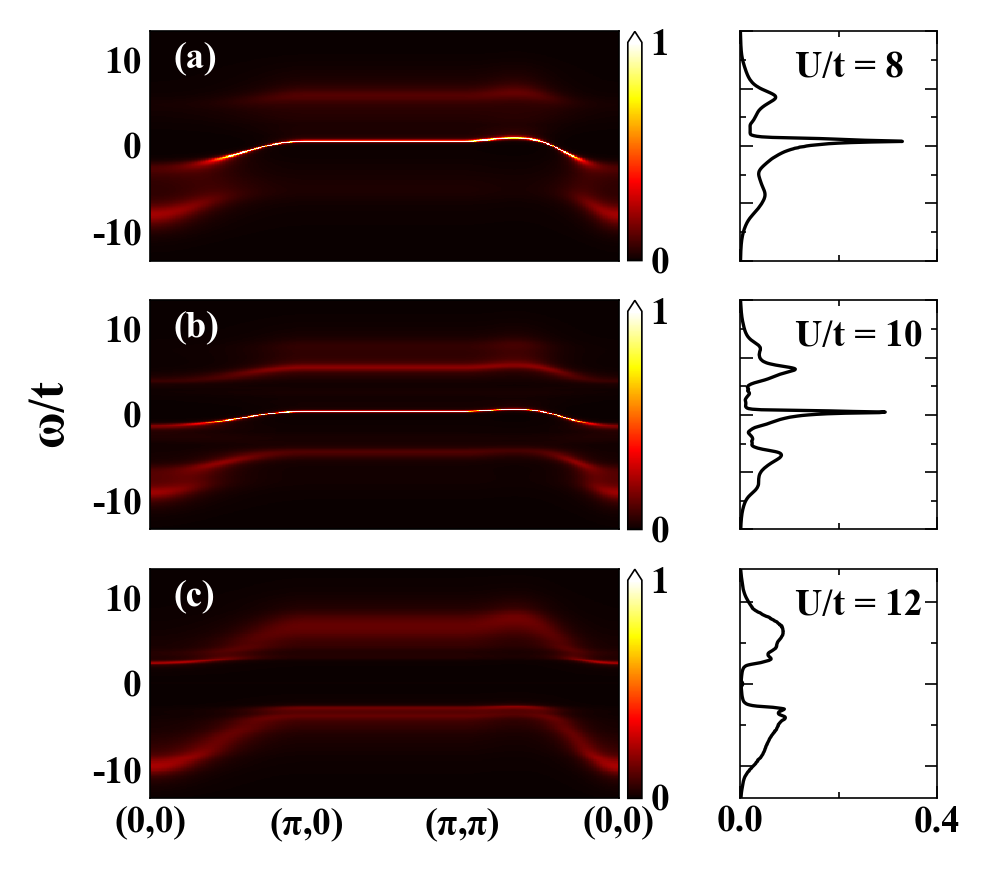}
\caption {\label{Fig4} (Color online) (Left) Electronic structures $A_{\sigma}^{\text{AC+ML}} (\omega,k)$ 
and (Right) $A_{\sigma}^{\text{AC+ML}} (\omega)$ for (a) $U/t=8.0$, (b) $10.0$, and (c) $12.0$. The results are determined by the 
AC+ML method of the converged QMC+DMFT self-energy $\Sigma{\sigma}^{\text{QMC+DMFT}} (i\omega_n)$ for the fully frustrated Hubbard 
model with $t'/t=1.0$ and $\mu=0.0$ on the square lattice.}
\end{figure}

We now present in Figs.~\ref{Fig4} (a)-(c) 
the electronic structure $A_{\sigma}^{\text{AC+ML}}
(\omega,k)$
and density of states $A_{\sigma}^{\text{AC+ML}} (\omega)$ for several $U/t$
values.
While the van Hove singularities are clearly seen in the metallic states with $U/t=8.0$, as in the case of the non-interacting system, 
the Hubbard bands emerging from electronic correlations are vaguely present around $\omega/t=\pm 5.0$. As $U/t$ 
increases, the position of the van Hove singularity  
moves towards  $\omega=0.0$ and there is a transfer of spectral weight towards the upper and lower Hubbard band.
At $U/t=12.0$ the system is a Mott insulator.


{\it Conclusions.-} We have proposed a novel analytic continuation
AC+ML technique in 
combination with IPT+DMFT that provides 
the self-energy $\Sigma{\sigma}^{\text{QMC+DMFT}} (\omega)$ from
 $\Sigma{\sigma}^{\text{QMC+DMFT}} (i\omega_n)$.
 The trained parameters for ML 
are determined from input and output training datasets
calculated simultaneously by IPT+DMFT  on Matsubara and 
real frequency domains, respectively.
The self-energy on the real frequency is obtained from the
ML process with (noisy) input QMC+DMFT self-energies in the Matsubara
domain and the trained ML parameters based on the convolutional autoencoder approach in each IPT+DMFT self-consistent step. When 
IPT+DMFT self-consistency including the ML process is completely satisfied, 
the patterns of QMC+DMFT self-energy on real frequency are rearranged. 
We demonstrated the powerfulness of the method for the case of the fully frustrated Hubbard model on
the square lattice where the QMC+DMFT data in the Matsubara domain are very noisy, and with the proposed method 
we were able to obtain trustable results at real frequencies.

Even though our AC+ML approach is computationally more expensive than
the maxEn method, it is not only free from bias from selections of ML 
training datasets and fitting parameters present in the maxEn method, 
but it directly performs analytic continuation of 
$\Sigma{\sigma}^{\text{QMC+DMFT}} (i\omega_n)$ as well,
which can be compared with experimental results. 
Therefore, we believe that our 
AC+ML method will be useful to investigate
 electronic properties of correlated systems which require
 reliable estimates of self-energies and spectral functions in 
 the real frequency domain.

\section{Acknowledgements}
We would like to thank H. Yoon and M. Han for fruitful discuss of ML process. This work was supported by the Ministry of Science 
through NRF-2018R1D1A1B07048139 (Hunpyo Lee), and Ministry of Education, Science and Technology 
NRF-2020R1I1A1A01071535 and POSTECH (Taegeun Song). RV acknowlegdes support
by the Deutsche Forschungsgemeinschaft (DFG, German Research Foundation) through TRR 288 - 422213477 (project B05).

\end{document}